\documentclass[aps,prl,twocolumn,showpacs,superscriptaddress,
groupedaddress]{revtex4}  

\usepackage{graphicx}  
\usepackage{dcolumn}   
\usepackage{bm}        
\usepackage{amssymb}   

\begin{document}
\DeclareGraphicsExtensions{.jpg,.pdf,.mps,.png,} 
\title{
{Quasi-relativistic calculus of graphene monolayer minimal
conductivity} }

\author{Halina~V.~Grushevskaya}
\email{grushevskaja@bsu.by}
\affiliation{Physics Department, Belarusan State University,
4 Nezalezhnasti Ave., 220030 Minsk, BELARUS}
\author{George~Krylov}
\email{krylov@bsu.by}
\affiliation{Physics Department, Belarusan State University,
4 Nezalezhnasti Ave., 220030 Minsk, BELARUS}

\begin{abstract}
We introduce a  quasi-relativistic theory of quantum transport in
graphene monolayer. It is based on the Dirac -- Hartry -- Fock
self-consistent field approximation, assumption on lattice
anti-ferromagnetic ordering  and an approach
[Falkovsky and Varlamov, Eur.~Phys.~J. {\bf B 56}, 281(2007)].
Minimal conductivity of graphene is shown to be  $4.83$ (in units
of $e^2/h$) when accounting for non-relativistic current only.
Allowing for quasi-relativistic corrections to current due to
process of pairs production and magneto-electric effects we obtain
the results for the minimal conductivity which are in a very good
agreement with experimental data for different supports.
\end{abstract}
\

 \maketitle
Today, graphene is considered as one of the most promising
materials for nanoelectronics. The ongoing boom in experimental
researches is not accompanied, however, by a substantial growth in
the number of theoretical papers, most of which are hitherto
appealed to sufficiently long enough proposed model  of
two-dimensional (2D) massless pseudo-Dirac fermions
\cite{Semenoff1984,Novoselov2005}. The last, despite the large
number of successful applications to the description of electronic
and optical properties of graphene under different conditions
(see, e.g. reviews \cite{Peres2009,Neto2009,Hancock2011}), can not
be considered as a final theory, because the model is a
nonrelativistic one. Theoretical estimates of monolayer graphene
minimal conductivity $\sigma_{min}$
\cite{Peres2009,Ziegler,Falkovsky,Ando2002} are in significant
discrepancies with experimental data
\cite{Novoselov2004,Dean,Bolotin,Du}.
So, the quantum-statistical theory of graphene with pseudo-Dirac
Hamiltonian gives $\sigma_{min}=\pi/2$ \cite{Falkovsky}. Here and
below in this paragraph, $ \sigma_{min}$ is measured in units of $
e^2/h$. At the same time, for example, for devices with large area
of graphene monolayer on SiO$_2$, the minimal conductivity of
graphene at low temperatures ($ \sim 1.5 $~K) turned out to be
$\sigma_{min} \sim 4 $ \cite{Novoselov2004}, and for graphene on
boron nitride substrate  $ \sigma_{min} \sim 6 $ \cite{Dean}.
Minimal conductivity of the suspended graphene
$\sigma_{min}\sim 6 $ at $T\sim 300 $~K according to \cite{Bolotin,Du}.

In papers \cite{myNPCS2013,we-arxiv,we-Kazan} a new approach has
been proposed to describe graphene electronic properties. It
utilizes the quasi-relativistic
 Dirac -- Hartry -- Fock self-consistent field
 approximation and assumption on anti-ferromagnetic
 ordering of the sublattices $ A, \ B $.
In this approach 2D graphene is described by the following
stationary equation for a second quantized fermion field $\widehat
{\tilde \chi }^{\dagger}_{+\sigma_{_B} }$:
\begin{eqnarray}
\left[ c
 \vec \sigma _{2D}^{BA} \cdot \vec p_{AB}
-\widetilde {\Sigma_{AB}\Sigma_{BA}} (\vec p) \right]\widehat
{\tilde \chi }^{\dagger}_{+\sigma_{_B} } (\vec r)\left|0,
\sigma
\right> \nonumber \\
= c E_{qu}(p) \widehat {\tilde \chi
}^{\dagger}_{+\sigma_{_B} } (\vec r) \left|0,
\sigma \right>
\label{rel-from-pseudi-Dirac-whithout-mix2}
\end{eqnarray}
where $ \widetilde {\Sigma_{AB} \Sigma_{BA}} =-\left(
\Sigma_{rel}^{x}\right)_{BA} \, \left(
\Sigma_{rel}^{x}\right)_{AB}$, $\vec \sigma _{2D}^{AB}= \left(
\Sigma_{rel}^{x}\right)_{BA} \vec \sigma _{2D} \left(
\Sigma_{rel}^{x}\right)_{BA}^{-1}$, $\vec \sigma _{2D}$ is the 2D
vector of the Pauli matrices,
$\vec p_{BA}
=
\left( \Sigma_{rel}^{x}\right)_{BA} \vec p\,
\left( \Sigma_{rel}^{x}\right)_{BA}^{-1}
$, $\vec p$ is the momentum operator,
$\widehat {\tilde \chi }^{\dagger}_{+\sigma_{_B}}  (\vec r)
\left|0,\sigma \right>= \left( \Sigma_{rel}^{x}\right)_{AB}
\widehat {\chi }^{\dagger}_{+\sigma_{_B} } (\vec r) \left|0,\sigma
\right>$, $c$ is the light speed,
2D transformation matrices  $\left( \Sigma_{rel}^{x}\right)_{BA},
\ \left( \Sigma_{rel}^{x}\right)_{AB}$ are determined by an
exchange interaction term $\Sigma_{rel}^{x}$ as
\begin{widetext}
\begin{eqnarray}
&\Sigma_{rel}^{x}\left(
\begin{array}{c}
\widehat {\chi } ^{\dagger}_{_{-\sigma_{_A}} }(\vec r) \\
\widehat {\chi }^\dagger _{\sigma_{_B}}(\vec r)
\end{array}
\right)\left|0,-\sigma \right> \left|0,\sigma \right>
=
 \left(
\begin{array}{cc}
0& \left( \Sigma_{rel}^{x}\right)_{AB}
\\
\left( \Sigma_{rel}^{x}\right)_{BA} & 0
\end{array}
\right)
\left(
\begin{array}{c}
\widehat {\chi }^{\dagger}_{-\sigma_{_A} } (\vec r) \\
\widehat {\chi} ^\dagger _{\sigma_{_B}}(\vec r)
\end{array}
\right)\left|0,-\sigma \right> \left|0,\sigma \right> \label{exchange}
, \\[3mm]
&\left( \Sigma_{rel}^{x}\right)_{AB}
\widehat {\chi }^\dagger _{\sigma_{_B}}(\vec r)\left|0,\sigma \right>
=
\sum_{i=1}^{N_v\,N}\int { d \vec r_i}
\widehat {\chi }^\dagger _{\sigma_i{^B}}(\vec r)\left|0,\sigma \right>
\langle 0,-\sigma_i|{\widehat \chi}^\dag_{-\sigma_i^A} (\vec r_i)
V(\vec r_i -\vec r)
{\widehat \chi}_{-\sigma_B}(\vec r_i)|0,-\sigma_{i'}\rangle ,
\label{Sigma-AB}
\\[3mm]
& \left( \Sigma_{rel}^{x}\right)_{BA}
\widehat {\chi }^{\dagger}_{_{-\sigma_{_A}} } (\vec r)
\left|0,-\sigma \right>
=\sum_{i'=1}^{N_v\,N}\int { d \vec r_{i'}}
\widehat {\chi }^{\dagger}_{_{-\sigma_{i'}^A} } (\vec r)
\left|0,-\sigma \right>
\langle 0,\sigma_{i'}|{\widehat \chi}^\dag_{\sigma_{i'}^B} (\vec r_{i'})
V(\vec r_{i'} -\vec r)
{\widehat \chi}_{_{\sigma_A}}(\vec r_{i'})|0,\sigma_i\rangle.
\label{Sigma-BA}
\end{eqnarray}
\end{widetext}
Due to the fact that
$\left( \Sigma_{rel}^{x}\right)_{BA}\neq
\left( \Sigma_{rel}^{x}\right)_{AB}$,
the vector $\vec p_{BA}$ of the Dirac cone 
 is somehow rotated and stretched in respect to the vector $\vec p_{AB}$ of its replica.
Energy dispersion law for 2D graphene  in the vicinity of the
Dirac points is also linear in this model, however the approach
allows to evaluate  cyclotron masses and charged carriers
asymmetry \cite{myNPCS2013}.
%
%

In this Letter we present a quasi-relativistic theory of quantum
transport in graphene monolayer and estimate its minimal
conductivity taking into account quasi-relativistic corrections
due to the process of pairs production ("Zitterbewegung"\
phenomenon) and magneto-electric effects (magnetotransport) for
$\mathrm{SiO}_\mathrm{2}$ and boron nitride supports, and for
suspended graphene.

Equation with electron-phonon interaction in graphene can be
obtained from (\ref{rel-from-pseudi-Dirac-whithout-mix2}) by
transition from $\vec p $ to generalized momentum ${\vec p} - e{\vec
A}/c $ as
\begin{eqnarray}
\left[ c \vec \sigma _{2D}^{BA} \cdot \left(\vec p_{AB} -{e\over
c}\vec A \right)-\widetilde {\Sigma_{AB}\Sigma_{BA}} \left(\vec
p_{AB}- e\vec A/c \right) \right]
\nonumber \\
\times \widehat {\tilde \chi
^{\dagger}_{+\sigma_{_B} }} (\vec r)\left|0,
\sigma \right>
= c E_{qu}(p) \widehat {\tilde \chi
^{\dagger}_{+\sigma_{_B} }} (\vec r) \left|0,
\sigma \right> .
\label{rel-from-pseudi-Dirac-whithout-mix3}
\end{eqnarray}
In subsequent,  we omit the sign  "\ $\!\widetilde{}$~"\ over
$\chi$. In (\ref{rel-from-pseudi-Dirac-whithout-mix3}) we expand
into a series the term $\widetilde {\Sigma_{AB}\Sigma_{BA}}$
depending on momentum, up to quadratic  terms inclusively:
\begin{widetext}
\begin{eqnarray}
\widetilde {\Sigma_{AB}\Sigma_{BA}}  (\vec p_{AB} -e\vec
A/c) =  \widetilde {\Sigma_{AB}\Sigma_{BA}}(0)+ \sum_i
\left. {d\widetilde {\Sigma_{AB} \Sigma_{BA}}\over d   p_i'}
\right|_{p_i'=0} \left( p_i ^{AB}
-{e\over c} A_i \right) %
\nonumber \\
+ {1\over 2} \sum_{i,j} \left. {d^2\widetilde {\Sigma_{AB}
\Sigma_{BA}}\over d   p_i' d   p_j'} \right|_{p_i', \ p_j' =0}
\left( p_i ^{AB} -{e\over c} A_i \right)\, \left( p_j ^{AB}
-{e\over c} A_j \right) + \ldots . \label{mass-renormalization2}
\end{eqnarray}
\end{widetext}
Appearing in the expression (\ref{mass-renormalization2}) matrices
are calculated in  $\pi(\mathrm{p}_z)$-electrons approximation
\cite{we-arxiv} in the corners $ K_A \ (K_B) $ of the Dirac cone
resulting to:
\begin{eqnarray}
\Sigma_{AB}(0)\Sigma_{BA}(0)=\left(
\begin{array}{cc}
 -0.058 & 0.015 \\
 -0.047 & 0.0079
\end{array}
\right),
\\ 
\left.\sum_{i=1}^{2}\frac{\partial^2}{\partial
{p_i}^2}\Sigma_{AB}\Sigma_{BA}
\right|_{K_A}=
\left(
\begin{array}{cc}
 -0.17 & 0.028 \\
 -0.15 & 0.021
\end{array}
\right),
\\ 
\left.\frac{\partial^2 \ \Sigma_{AB}\Sigma_{BA}}{\partial {p_x}\partial
{p_y}}\right|_{K_A}= \left(
\begin{array}{cc}
 -11.0+6.2 i & 1.7 -0.95 i \\
 -8.2+4.6 i & 0.40 -0.34 i
\end{array}
\right). \label{mix-deriv}
\end{eqnarray}

Quasi-relativistic expression \cite{Davydov} for the current
operator    $ j_i^{G}$ for graphene reads
\begin{eqnarray}
j_i^{G} =\frac{j_i}{c}=c^{-1}\left( j_i^{O}+j_i^{Z}+j_i^{SO}\right),
 \label{graphene-quasirel-current2}
\\
j_i^{O}=e \chi^{\dagger}_{+\sigma_{_B} } (x^+) v^i_{x^+x^-}
\chi_{+\sigma_{_B} } (x^-),\\
j_i^{Z}= -{e^2 A_i\over c\,
 \widetilde {\Sigma_{AB}\Sigma_{BA}}(0)}
 \chi_{+\sigma_{_B} }^\dagger  \chi_{+\sigma_{_B} },
\\
j_{2(1)}^{SO} =
 (-1)^{1(2)} {\imath e \over 2 }
v^{1(2)}_{x^+ x^-}\chi_{+\sigma_{_B} }^\dagger  \sigma_z
 \chi_{+\sigma_{_B} }
 .
\end{eqnarray}
Here
\begin{eqnarray}
x^{\pm}=x \pm\epsilon ,\ x=\{\vec r,\ t_0 \},\ t_0=0,\ \epsilon
\to 0 ;\label{current-limits}
\end{eqnarray}
$\vec v $ is the velocity operator defined by the momentum
derivative of the Hamiltonian
(\ref{rel-from-pseudi-Dirac-whithout-mix3}).
The terms $j_i^{O},\ j_i^{Z},\ j_i^{SO}$ in
(\ref{graphene-quasirel-current2}) describe, respectively, the
current component that satisfies Ohm's law and  contributions of
polarization and magnetoelectric effects.
In the interaction representation, the current is expressed through
the evolution operator $ \hat U(x'-x'') $ according to the formula
\begin{widetext}
\begin{eqnarray}\label{funct-represent}
\chi^{\dagger}_{+\sigma_{_B} } (x^+) v^l_{x^+x^-}
\chi_{+\sigma_{_B} } (x^-) =\hat
U(x^{+}-x')\chi^{\dagger}_{+\sigma_{_B} } (x') v^l_{x'\, x^-} \hat
U(x^{-} - x'')\chi_{+\sigma_{_B} } (x'')
\nonumber \\
=
\left[1- {\imath 
}\int V^{G}(x^{+}-x') dt' d\vec {x'} +\ldots \right]
\chi^{\dagger}_{+\sigma_{_B} } (x')v^l_{x'x^-} \chi_{+\sigma_{_B}
} (x^-)
\nonumber \\
=\left\{1-
{(-\imath) 
}\int \int \chi^{\dagger}_{+\sigma_{_B} } (x^{+}-x') \sum_i \left[
{e\over c}  v^i_{x^{+}-x',\, \bar{x}}
 A_i(\bar{x})
+\widetilde {\Sigma_{AB}\Sigma_{BA}}(0) {d\widetilde {\Sigma_{AB}
\Sigma_{BA}}\over d   p_i^{AB}} (0) \ {v^i_{x^{+}-x',\, \bar{x}}
}\right. \right. 
+  {\widetilde {\Sigma_{AB}\Sigma_{BA}}(0)^2\over 2} \\ \nonumber
\left.\left.\times\sum_{j} {d^2\widetilde {\Sigma_{AB} \Sigma_{BA}}\over d
p_i^{AB} d   p_j^{AB}} (0)\ {v^i_{x^{+}-x',\, \bar{x}}\,
v^j_{x^{+}-x',\ \bar{x}}
 }
\right]\chi_{+\sigma_{_B} } (\bar{x})
   d\bar{t}\, d\vec {\bar{x}}\, dt'\, d\vec {x'}
+ \ldots \right\} 
\chi^{\dagger}_{+\sigma_{_B} } (x') v^l_{x'x^-} \hat
U(x^{-}-x'')\chi_{+\sigma_{_B} } (x'')
\end{eqnarray}
\end{widetext}
where $ V ^ {G} $ is the interaction between the electromagnetic
field $ \vec A $ and the current $ \chi ^ {\dagger} _ {+ \sigma_
{_B}} (x ') v ^ i_ {x'x} \chi_ {+ \sigma_ {_B}} (x) $. Using the
relationship $ U = \imath G_1 $ of the evolution operator with the
one-particle Green function $ \imath G_1 (x - x') =
\chi^{\dagger} (x)\chi (x') $ \cite{Krylova monography2} and
accounting for (\ref{funct-represent}), we rewrite, for example,
$j_i^{O}$ in (\ref{graphene-quasirel-current2}) through  $G_1$ and
two-particle Green function $ G_2(x,\, x',\, \bar{x},\,
\bar{x}') =(\imath)^2 \chi^{\dagger} (x)\chi^{\dagger} (x')\chi
(\bar{x})\chi (\bar{x}')$:
\begin{widetext}
\begin{eqnarray}
j_l^{O}
={e\over (\imath)^3} \left\{1-
{(-\imath)
}\int \int \sum_i \left[ {e\over c}  v^i_{x^{+}-x',\, \bar{ x}}
 A_i(\bar{x})
+\widetilde {\Sigma_{AB}\Sigma_{BA}}(0) {d\widetilde {\Sigma_{AB}
\Sigma_{BA}}\over d   p_i^{AB}} (0) \ {v^i_{x^{+}-x',\, \bar{ x}}
}\right. \right.
+  {\widetilde {\Sigma_{AB}\Sigma_{BA}}(0)^2\over 2}\nonumber
\\
\left.\left. \times
\sum_{j} {d^2\widetilde {\Sigma_{AB} \Sigma_{BA}}\over d
p_i^{AB} d   p_j^{AB}} (0)\ {v^i_{x^{+}-x',\, \bar{ x}}\,
v^j_{x^{+}-x',\ \bar{ x}}
 }
\right]\right.
\left. G_2(x^{+},\, \bar{x},\, x',\, x'') G_1( x^{-} -x'')
d\bar{t}\, d^2 \vec {\bar{ x}}\, dt'\, d^2\vec {x'}\, dx'' +
\ldots \right\} v^l_{x'x^{-}} .\quad \ \
\label{green--representation1}
\end{eqnarray}
Let us turn to the imaginary time, perform the Fourier -- Laplace
transform of the expression (\ref{green--representation1}), and
obtain
\begin{eqnarray}
j_i^{O}(\omega, \ k)= \mbox{Tr }\ {e^2\over c (2\pi)^2} {1\over
(\imath)^3}
 \int
  v^i_{(\vec {p}^+ -\vec {p}^-)}
 A_i
 \left[
- \imath \bar{\beta} {f[\beta ((H(p^-)-\mu)/\hbar)] - f[\beta
((-H(p^+)-\mu)/\hbar)]\over \beta ( z^+ - z^-) -
\beta(H(p^-)/\hbar) - \beta(H(p+)/\hbar) } \right]
 v^i_{(\vec p^+ - \vec p^-)}
\nonumber \\
\times {1\over -\imath(\omega -\hat\omega ^+ + \hat \omega^ - )}
\delta[\beta(z^- - z^+) - \beta(z)] \,   d\vec {p}  \, dz^- .\ \ \
 \label{Laplas-Fourier-transform6}
\end{eqnarray}
\end{widetext}
Here the Hamiltonian $H$ satisfies $H(p)\, \chi_p(\vec r)= E(p)\,
\chi_p(\vec r)$, $\hat\omega^{+}= H^{\dagger}(-p^{+})/\hbar $,
$\hat\omega^{-}= H(p^{-})/\hbar $ owing to
$\imath \hbar {\partial \over \partial t^{\pm}} \chi_p(\vec r,\,
t^{\pm}) = H(p^{\pm})\,
 e^{\imath \omega^{\pm}t^{\pm}}\chi_p(\vec r)$, $t^+> 0$, $t^- < 0$;
$\bar{\beta}=1/T$, $\beta(z)=\bar{\beta}\hbar z$.
A coefficient at $ A_i $ entering into the expression
(\ref{Laplas-Fourier-transform6}), after its division by $ c $,
gives the ohmic contribution to the graphene conductivity:
\begin{widetext}
\begin{eqnarray}
\sigma_{ii}^{O}(\omega, \ k)=
 {\imath e^2\over c^2 (2\pi)^2} {\bar{\beta  }}^2
 \int
\left(
 \ \mbox{Tr}\
  \vec v\,^i_{(\vec {p}^+ -\vec {p}^-)}
 M\, , \,
\vec v\, ^i_{(\vec p^+ - \vec p^-)}
N \right)\,   d
\vec {p}
,\ \ \
 \label{conduction3}
\end{eqnarray}
where $(\cdot\, , \, \cdot) $ is the scalar product,
$i=1, 2$; operators $ M,\  N$ are defined as
\begin{eqnarray}
M=
 {f[\bar{\beta} (H(p^-)-\mu)]
- f[\bar{\beta} (-H(p^+)-\mu)]\over \bar{\beta} \hbar z -
\bar{\beta}H(p^-) - \bar{\beta} H(p+) }, \quad N= {1\over (\hbar
\omega - H^\dagger (- p^+) + H(p^-))\bar{\beta}}.
\label{operatorsM&N}
\end{eqnarray}
Analogous  calculations yield
\begin{eqnarray}
 \sigma_{ll}^{Z}(\omega, \ k)=
\left(\mbox{Tr }\ {-\imath e^2\over c^2 (2\pi)^2}
{\bar{\beta }}^2
 {\widetilde {\Sigma_{AB}\Sigma_{BA}}(0)\over 2}\sum_{i=1}^2
{d^2\widetilde {\Sigma_{AB} \Sigma_{BA}}\over d   p_i^{2}} 
\int
  \vec v\,^i_{(\vec {p}^+ -\vec {p}^-)}
M
 \, , \, \vec v\,^i_{(\vec p^+ - \vec p^-)}
N\right) \,   d
\vec {p}
   ,\ \ \
 \label{Zitterbewegung_conduction1}\\
 \sigma_{22(11)}^{SO}(\omega, \ k, \sigma_z)=
(-1)^{1(2)} {\imath \over 2 } \left(
\mbox{Tr }\ {-\imath e^2\over c^2 (2\pi)^2} {\bar{\beta }}^2
 {\widetilde {\Sigma_{AB}\Sigma_{BA}}(0)\over 2}
{d^2\widetilde {\Sigma_{AB} \Sigma_{BA}}\over d p_{1(2)} d
p_{2(1)}}
(0)
\int
 \vec v\, ^{1(2)}_{(\vec {p}^+ -\vec {p}^-)}
M\sigma_z \, , \, \vec v\, ^{1(2)}_{(\vec p^+ - \vec p^-)}
N \right)\,   d
\vec {p}
.
 \label{spin-orbit-conduction1}
\end{eqnarray}
\end{widetext}

We utilize the following approximation:
\begin{eqnarray}
\vec v_{AB}\approx {\partial H_0\over  \partial \vec p}
\label{velocity-in-diagonal-represenation_0}
\end{eqnarray}
corresponding to massless case. Here $ H_0 = c \vec \sigma _ {2D}
^ {BA} \cdot \vec p_ {AB} $ is the unperturbed electron
Hamiltonian of the problem
(\ref{rel-from-pseudi-Dirac-whithout-mix3}). We denote eigenvalues
of the unperturbed electron Hamiltonian $ H_0 $ by $ E^e_{1, \, 2}
$,  eigenvalues of the hole Hamiltonian $H_0^{\dagger} $ by $
E^h_{1, \, 2} $.
In representation where Hamiltonian $H_0$ is diagonal,
 taking the trace operation
in (\ref{conduction3}, \ref{Zitterbewegung_conduction1},
\ref{spin-orbit-conduction1}) can be easily carried out.
We make a change: $-H(p) \to H^\dagger (-p) $ and
for each band $ a $, $ a = 1, \, 2$  introduce the  Hamiltonians
of quasiparticles $ H_0^{a}$, $ \mathop {H_0^{a}}^\dagger$ with
the eigenvalues $E^e_{a}$, $E^h_{a}$ to quantize $M$ and $N$
(\ref{operatorsM&N}):
\begin{widetext}
\begin{eqnarray}
M =
 {f[\bar{\beta} (H(p^-)-\mu)]
- f[\bar{\beta} ( H^\dagger (-p^+)-\mu)]\over \bar{\beta} \hbar z
- \bar{\beta}H(p^-) +\bar{\beta} H^\dagger (-p^+) }= \left\{M_{ab}
\right\},
\ \ \ M_{ab}=
 {f[\bar{\beta} (H_0^{a}(p^-)-\mu)]
- f[\bar{\beta}( \mathop{H_0^{b}}^\dagger (-p^+)-\mu)]\over
\bar{\beta} \hbar z - \bar{\beta}H_0^{a}(p^-) + \bar{\beta}
\mathop{H_0^{b}}^\dagger (-p^+) };\quad
\label{matrix-quantization2}\\
N= {1\over (\hbar \omega - H^\dagger (-p^+) + H(p^-))\bar{\beta}}=
\left\{ N_{ab}\right\},\ \ N_{ab} = {1\over (\hbar \omega -
\mathop{H_0^a}^\dagger (-p^+) + H_0^b(p^-))\bar{\beta}} .
\label{matrix-quantization3}
\end{eqnarray}
\end{widetext}

For a degenerate Dirac cone, we have
 $E^e_{1,\, 2}(p)\approx \mp c v_F \hbar p$,
$E^h_{1,\, 2}(p)\approx \pm c v_F \hbar p$ where $v_F$ is the
Fermi velocity. In this case, we can calculate (similar to
\cite{Falkovsky})  the non-relativistic contribution $ \sigma_{aa,
\,ij} ^ {O} $ to graphene conductivity stipulated by transitions
of quasiparticles between states of the same band,  and get
\begin{widetext}
\begin{eqnarray}
\sigma_{aa,\, ij}^{O}(\omega, \ k)=
 {\imath e^2\over c^2 (2\pi)^2}
 \int
{f[\bar{\beta} (H_0^{a}(p^-)-\mu)] - f[\bar{\beta}(
\mathop{H_0^{a}}^\dagger (-p^+)-\mu)]\over  \hbar z -
 H_0^{a}(p^-) +   \mathop{H_0^{a}}^\dagger
(-p^+) }
{v^{i}_{aa}(p)v^{j}_{aa} (p)\ d\vec {p}\over
(\hbar \omega - \mathop{H_0^a}^\dagger (-p^+) +
H_0^a(p^-)) }
\nonumber \\
= -
{\imath e^2\over c^2  2\pi h} 
\int
{\omega(k)v^{i}_{aa}(p)  v^{j}_{aa} (p)
\partial f[\bar{\beta} (E_{a}(p)-\mu)]/\partial E_a(p)
\over
  z  \omega -\omega(k) \omega
+  z \omega(k) - \omega^2(k)}\ d\vec {p}, \quad  i, j\in
\{x,y\}.
 \label{intra-zone-conduction}
\end{eqnarray}
\end{widetext}
The last formula differs from that of \cite{Falkovsky} only by the
angular dependence of the integrand due to difference for the
velocity operators in the models.

Since the matrix of mixed-derivatives (\ref{mix-deriv}) is
complex, magneto-transport contribution
(\ref{spin-orbit-conduction1}) to the conductivity will be
stipulated by both interband and intraband summands. For small $ k
$ and low temperatures  we use the limit $\tanh \omega/T\to 1$
\cite{Falkovsky}. In this limit, the intraband contribution is
zero, and for the real part of conductivity due to interband
transitions, we get the following values for separate
contributions: $\sigma^{O}=4.83e^2/h$, $\sigma^{Z}=0.74e^2/h$,
$\sigma^{SO}=\pm 0.41 e^2/h$. If the process of electron--hole
pairs production is broken, then
$\sigma=\sigma^{O}-|\sigma^{SO}|=4.42e^2/h$, that gives the
experimental value of conductivity for  graphene monolayer on
SiO$_2$. If the process of electron -- hole pairs production is
not broken, then $\sigma=\sigma^{O}+\sigma^{Z}+|\sigma^{SO}|=5.98
e^2/h$, and this is precisely the minimal conductivity of graphene
on boron nitride support  and of suspended graphene. There will be
also observable  trembling in magnetic fields:
$\sigma=\sigma^{O}+\sigma^{Z}+\sigma^{SO}$.

To summarize,  the quantum-statistical theory of graphene with
Dirac -- Hartry -- Fock Hamiltonian has been proposed which
assumes the lattice anti-ferromagnetic ordering  and takes into
account of polarization and magnetoelectric effects.
Calculation of minimal conductivity of graphene monolayer within this
theory gives the values recorded in the experiments for different supports.

\end{document}